\documentclass[a4paper]{report}
\usepackage[utf8]{inputenc}
\usepackage[T1]{fontenc}
\usepackage{RJournal}
\usepackage{amsmath,amssymb,array}
\usepackage{booktabs}

\begin{document}

\sectionhead{Contributed research article}
\volume{XX}
\volnumber{YY}
\year{20ZZ}
\month{AAAA}

\begin{article}
\title{Analysis of the Results of Metadynamics Simulations
by \pkg{metadynminer} and \pkg{metadynminer3d}}
\author{by Dalibor Trapl and Vojtěch Spiwok}

\maketitle

\abstract{
The molecular simulations solve the equation of motion of molecular systems,
making 3D shapes of molecules four-dimensional by adding the time coordinate.
These methods have a great potential in drug discovery because they
can realistically model the structures of protein molecules targeted by drugs as
well as the process of binding of a potential drug to its molecular target.
However, routine application of biomolecular simulations is hampered by the very
high computational costs of this method. Several methods have been developed
to address this problem. One of them, metadynamics, disfavors states of
the simulated system that have been already visited and thus forces
the system to explore new and new states. Here we present the package
\pkg{metadynminer} and \pkg{metadynminer3d} to analyze and visualize
results from metadynamics, in particular those produced by a popular
metadynamics package Plumed.
}

\section{Introduction}

Molecular simulations and their pioneers Martin Karplus, Michael Levitt, and Arieh Warshel
have been awarded Nobel Prize in 2013 \citep{nobel}. These methods, in particular the method
of molecular dynamics simulation, computationally simulate the motions of atoms in a molecular
system. A simulation starts from a molecular system defined by positions (Cartesian
coordinates) of individual atoms. The heart of the method is in calculation of forces acting
on individual atoms and their numerical integration in the spirit of
Newtonian dynamics, i.e., conversion of a force vector to acceleration vector,
velocity vector and, finally, to a new position of an atom. By repeating these steps,
it is possible to reconstruct a record of atomic motions known as a trajectory.

Molecular simulations have a great potential in drug discovery. A molecule of drug
influences (enhances or blocks) the function of some biomolecule in the patient's
body, typically a receptor, enzyme or other protein. These molecules are called
drug targets. The process of design of a new drug can be significantly accelerated by knowledge
of the 3D structure (Cartesian coordinates of atoms) of the target.
With such knowledge, it is possible to find a "druggable" cavity in
the target and a molecule that fits and favourably binds
to this cavity to influence its function.
Strong binding implies that the drug influences the target even in low doses, hence
does not cause side effects by interacting with unwanted targets.

Experimental determination of 3D structures of proteins and other biomolecules
is very expensive and laborious process. Molecular simulations can, at least
in principle, replace such expensive and laborious experiments by computing.
In principle, a molecular simulation starting from virtually any 3D shape of
a molecule would end up in energetically the most favourable shape. This is analogous
with water flowing from mountains to valleys and not in the opposite way.

Unfortunately, this approach is extremely computationally expensive.
The integration step of a simulation must be small enough to
comprise the fastest motions in the molecular system. In practical simulations, it is necessary
to use femtosecond integration steps. This means that it is necessary to carry out
thousands of steps to simulate picoseconds, millions of steps to simulate nanoseconds, and so
forth. In each step, it is necessary to evaluate a substantial number of interactions between atoms.
As the result of this, it is possible to routinely simulate nano- to microseconds.
Longer simulations require special high-performance computing resources.

Protein folding, i.e., the transition from a quasi-random to the biologically relevant
3D structure, takes place in microseconds for very small proteins and in much longer
time scales for pharmaceutically interesting proteins. For this reason, prediction
of a 3D structure by molecular simulations is limited to few small and fast folding
proteins. For large proteins it is currently impossible or at least far from being routine.

Several methods have been developed to address this problem. Metadynamics \citep{mtd}
uses artificial forces to force the system to explore states that have not been previously
explored in the simulation. At the beginning of the simulation, it is necessary to
chose some parameters of the system referred to as collective variables. For example,
numerically expressed compactness of the protein can be used as a collective variable
to accelerate its folding from a noncompact to a compact 3D structure.
Metadynamics starts as a usual simulation. After a certain number of steps
(typically 500), the values of the collective variables are calculated an from this moment
this state becomes slightly energetically disfavored due to the addition of
an artificial bias potential in the shape of a Gaussian hill.
After another 500 steps, another
hill is added to the bias potential and so forth. These Gaussian hills accumulate until
they "flood" some energy minimum and help the system to escape this minimum
and explore various other states
(Figure~\ref{figure:Metadynamics}). In analogy to water floating from mountains
to valleys, metadynamics adds "sand" to fill valleys to make water flow from
valleys back to mountains. This makes the simulation significantly more efficient
compared to a conventional simulation because "water" does not get stuck anywhere.

By the application of metadynamics, it is possible to significantly accelerate the process
of folding. Hopefully, by the end of metadynamics we can see folded, unfolded, and many
other states of the protein. However, the interpretation of the trajectory is not
straightforward. In standard molecular dynamics simulation (without metadynamics),
the state which is the most populated is the most populated in reality. This is
not true anymore with metadynamics. Here comes the time for \pkg{metadynminer} and
\pkg{metadynminer3d}.

\pkg{Metadynminer} and \pkg{metadynminer3d} use the results of metadynamics simulations
to calculate the free energy surface of the molecular system.
The most favoured states (states most populated in reality) correspond to minima
on the free energy surface. The state with the lowest free energy is the most
populated state in the reality, i.e., the folded 3D structure of the protein.

As an example to illustrate metadynamics and our package, we use
an ultrasimple molecule of "alanine dipeptide"
(Figure \ref{figure:Metadynamics}). This molecule can be viewed as a "protein" with
just one amino acid residue (real proteins have hundreds or thousands of amino acid
residues). As a collective variable it is possible to use an angle $\phi$
defined by four atoms. Biasing of this collective variable accelerates a slow
rotation around the corresponding bond. Figure \ref{figure:Metadynamics} shows the
free energy surface of alanine dipeptide as the black thick line. It is not known
before the simulation. The simulation starts from the state B. After 500 simulation
steps, the hill is added (the hill is depicted as the red line, the flooding potential ("sand")
at the top, the free energy surface with added flooding potential at the bottom).
Sum of 10, 100, 200, 500, and 700 hills is depicted as yellow to blue lines.

At the end of simulation the free energy surface is relatively well flattened
(blue line in Fig. \ref{figure:Metadynamics} bottom). Therefore, the free energy
surface can be estimated as a negative imprint of added "sand":

\begin{equation}
G(s) = -kT \log(P(s)) = -V(s) = \sum_i w_i \exp(-(s-S_i)^2/{2 \sigma^2})
\label{mtd}
\end{equation}

\noindent where $G$, $V$, and $P$ are free energy, metadynamics bias (flooding)
potential, and probability,
respectively, of a state with a collective variable $s$, $k$ is Boltzmann constant,
$T$ is temperature in Kelvins, $w_i$ is height, $S_i$ is position and $\sigma_i$
is width of each hill. The equation can be easily generalized for two or more CVs.

\begin{widefigure}[htbp]
  \centering
  \includegraphics{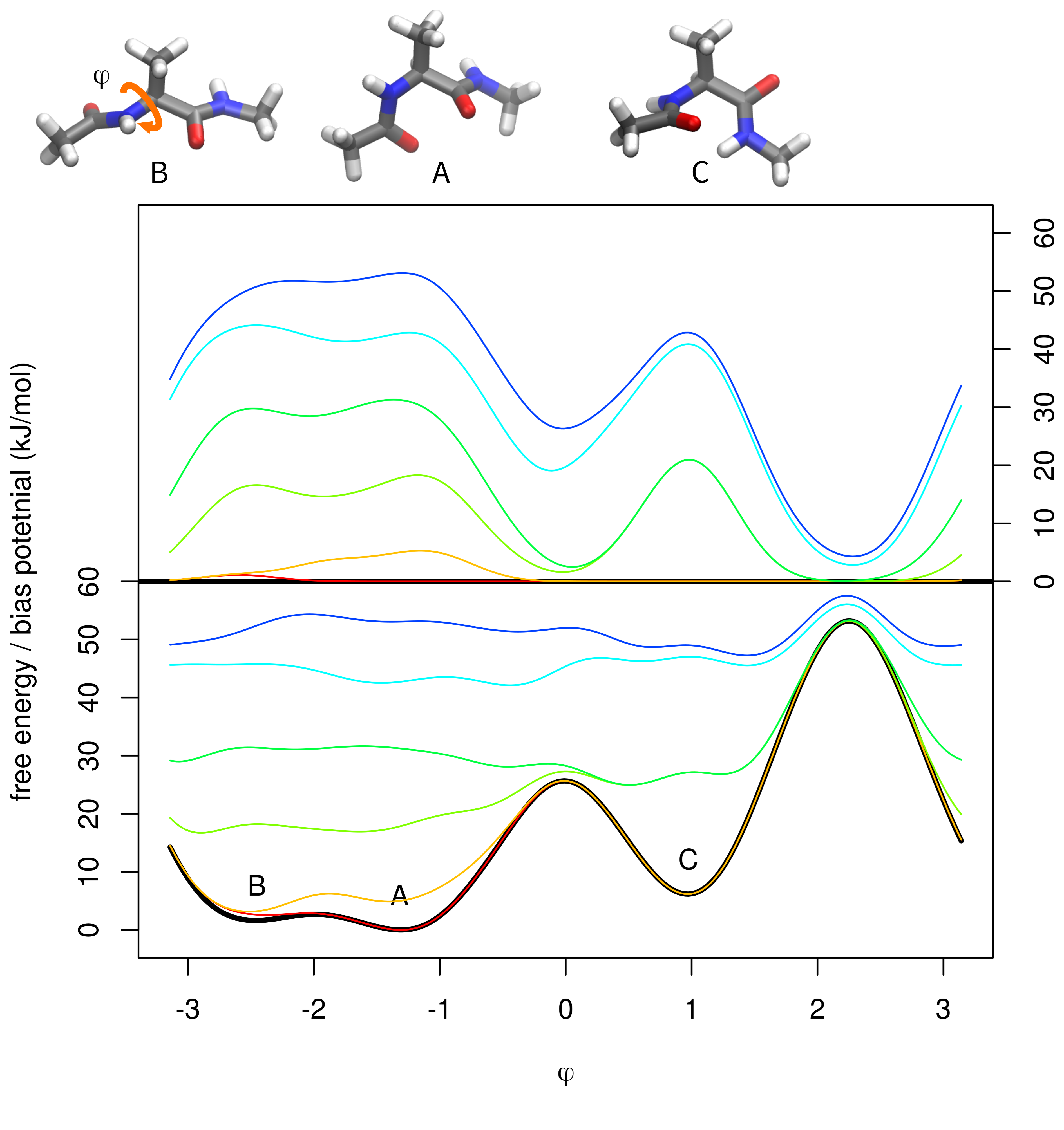}
  \caption{Metadynamics simulation of alanine dipeptide. Dihedral angle $\phi$ was used
           as the collective variable. The top part shows molecular structures of three
           free energy minima (stable structures) differing in the value of $\phi$.
           According to metadynamics prediction, A is the global minimum (free energy
           0 kJ/mol) and B and C are local minima (1.5 and 6.3 kJ/mol, respectively).
           According to Equation 1, this corresponds to probabilities
           0.61, 0.34, and 0.05 for A, B, and C, respectively. The middle part shows the
           bias potential (scaled by $(T+\Delta T)/\Delta T$) after addition of 1, 10,
           100, 200, 500, and 700 hills (colors from red to blue). The bottom part shows the
           accurate free energy surface calculated by metadynamics with 30,000 hills (black)
           flooded by 1, 10, 100, 200, 500, and 700 hills (colors from red to blue).
           The figure was generated by \pkg{metadynminer} except for molecular structures
           and final assembly.}
  \label{figure:Metadynamics}
\end{widefigure}

The original version of metadynamics was developed with constant heights of Gaussian hills.
Later, a so-called well-tempered metadynamics was developed \citep{wtmtd}, which uses
decreasing hill heights to improve the accuracy of the results. This requires modification
of the equation:

\begin{equation}
G(s) = -kT \log(P(s)) = - \frac{T + \Delta T}{\Delta T} V(s) =
- \frac{T + \Delta T}{\Delta T} \sum_i w_i \exp(-(s-S_i)^2/{2 \sigma^2})
\label{wtmtd}
\end{equation}

\noindent where $\Delta T$ an input parameter with the dimension of temperature
(zero for unbiased simulation and infinity for the original metadynamics with constant
hill heights). Nowadays, the vast majority of metadynamics applications use the
well-tempered metadynamics algorithm for its better convergence towards accurate
free energy surface prediction.

There are numerous packages for molecular simulations such as Amber \citep{amber},
Gromacs \citep{gmx}, Gromos \citep{gms}, NAMD \citep{namd},
CHARMM \citep{charmm}, Acemd \citep{acemd}, and others.
These packages are primarily developed for basic unbiased simulations with no or very
limited support of metadynamics. Plumed software \citep{plumed} has been developed to
introduce metadynamics into various simulation programs. Since its introduction,
Plumed articles have been cited in more than thousand papers from drug design,
molecular biology, material sciences, and other fields. The R package \pkg{metadynminer}
was developed for analysis and visualization of the results from Plumed. With
a simple file conversion script, it can be used also with other simulation
programs that support metadynamics.

\section{Example of usage}

\pkg{Metadynminer} will be presented on a bias potential from a
30 ns (30,000 hills) simulation of alanine dipeptide (Figure~\ref{figure:Metadynamics}).
Two rotatable bonds of the molecule, referred to as $\phi$ and $\psi$,
were used as collective variables. This is basically an expansion of the free energy
surface in Figure \ref{figure:Metadynamics} to two dimensions.
Hills from simulations with two collective variables ($\phi$ and $\psi$) and with
one collective variable ($\phi$) are provided in \pkg{metadynminer} as
\code{acealanme} and \code{acealanme1d}, respectively. \pkg{Metadynminer3d} was developed
for analysis of metadynamics with three collective variables. It contains a sample
data \code{acealanmed3}, with collective variables $\phi$, $\psi$ and $\omega$.
We decided to distribute \pkg{metadynminer} and \pkg{metadynminer3d} separately, because of
the use of different visualization tools and to keep the size of packages low.
Metadynamics simulations with 1-3 CVs comprise almost all metadynamics
applications nowadays (not considering special metadynamics variants).

Hills file generated by Plumed package (filename "HILLS") can be loaded to R by
the function \code{read.hills}:
\begin{example}
  hillsfile <- read.hills("HILLS", per=c(T, T))
\end{example}
The parameter \code{per} indicates periodicity of the collective variable (dihedral angles
are periodic, i.e., $+\pi \simeq -\pi$).

\begin{figure}[htbp]
  \centering
  \includegraphics{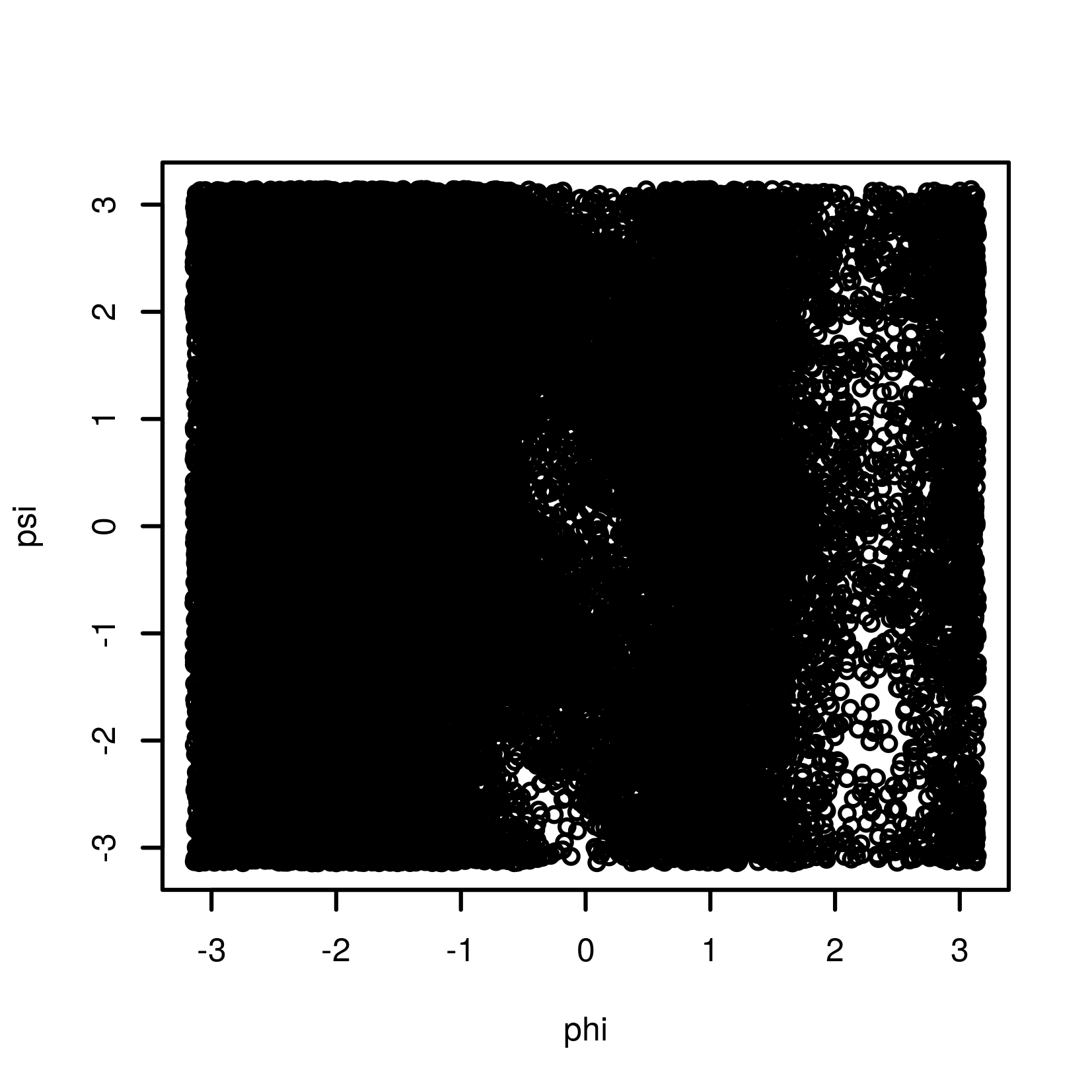}
  \caption{Scatter plot of hills position.}
  \label{figure:plothills}
\end{figure}

Typing the name \code{hillsfile} will return its dimensionality (the number of CVs) and
the number of hills. A hills object can be easily plotted:
\begin{example}
  plot(hillsfile, xlab="phi", ylab="psi")
\end{example}
For metadynamics with one collective variable, it plots its evolution. For metadynamics
with two or three collective variables, it plots a scatter plot of collective variables
number 1 vs. 2 or 1 vs. 2 vs. 3, respectively (Figure \ref{figure:plothills}).

\begin{figure}[htbp]
  \centering
  \includegraphics{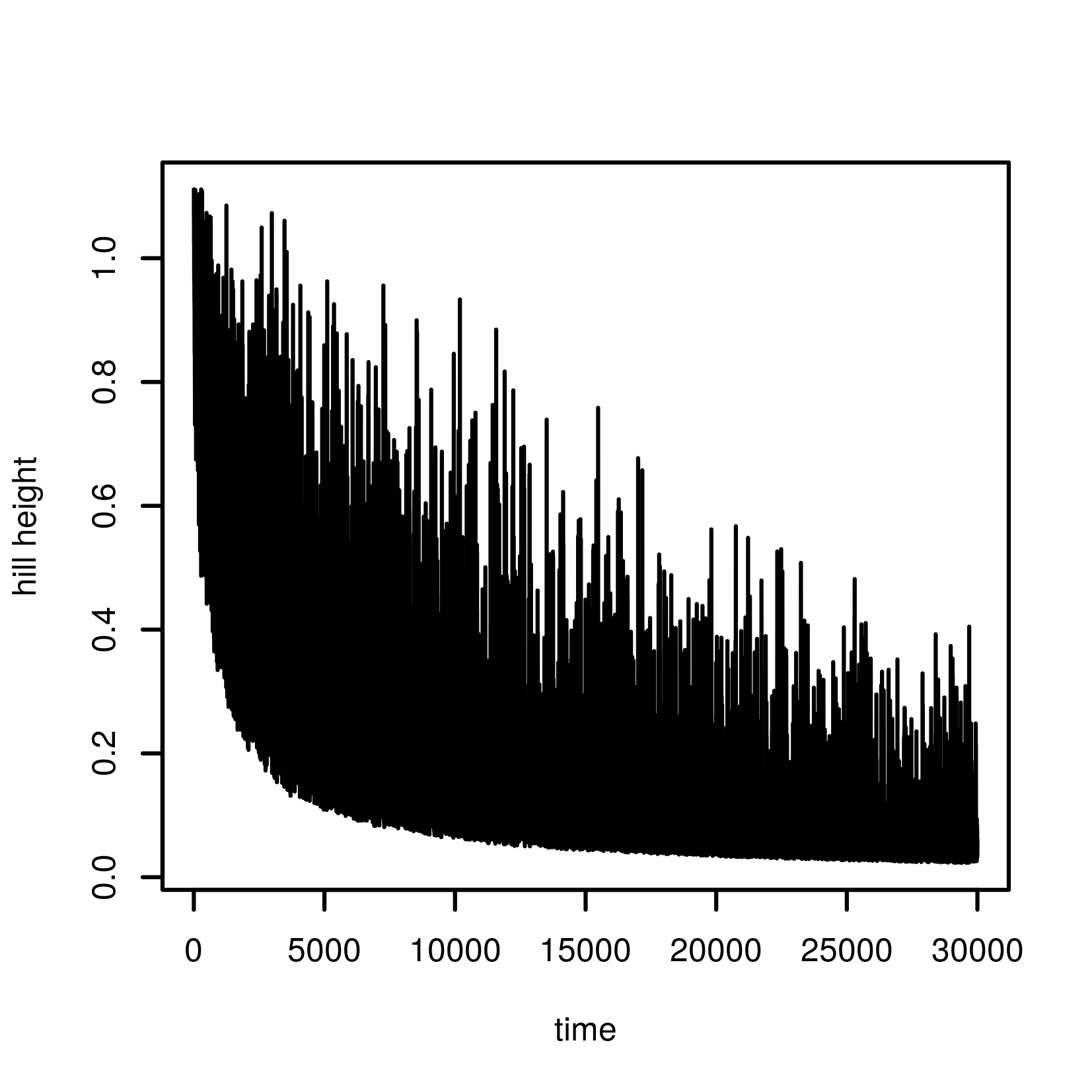}
  \caption{Evolution of heights of hills in metadynamics
           plotted by function \code{plotheights}.}
  \label{figure:plotheights}
\end{figure}

In well-tempered metadynamics it may be interesting to see the evolution
of hill heights ($w_i$ in Equation \ref{wtmtd}).
This can be plotted (Figure \ref{figure:plotheights}) by typing:
\begin{example}
  plotheights(hillsfile)
\end{example}

Addition operation is available for hillsfile object. For example, multiple hills
files can be concatenated.

Next, the user can sum negative values of all hills to make the free energy surface
estimate by typing:
\begin{example}
  fesurface <- fes(hillsfile)
\end{example}
Hills files from well-tempered metadynamics are prescaled by $(\Delta T + T)/\Delta T$
when printed by Plumed, so no special action is required in \pkg{metadynminer}.
The function \code{fes} uses the Bias Sum algorithm \citep{petr}. This function is
fast because instead of evaluation of Gaussian function for every hill, it uses
a precomputed Gaussian hill that is relocated to hill centers.
It is also fast because it was implemented in C++ via \CRANpkg{Rcpp}. Because of approximations
used in the function \code{fes}, this function should be used for visualization purposes.
For detailed analysis of a free energy surface, we advice to use a slow but accurate 
\code{fes2} function. This function explicitly evaluates Gaussian function for every
hill. It can be also used for (rarely used) metadynamics with variable hill widths.

\begin{figure}[htbp]
  \centering
  \includegraphics{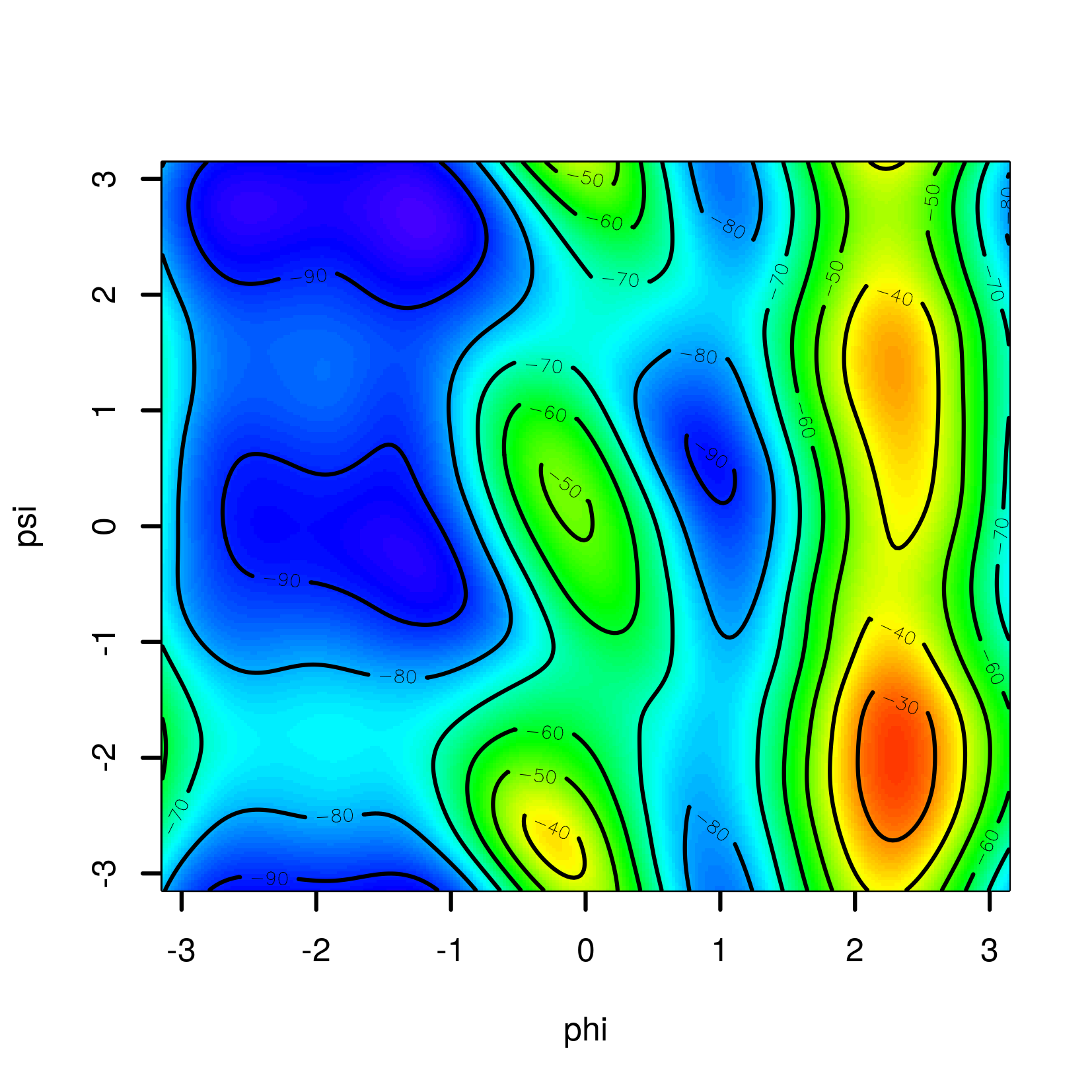}
  \caption{Free energy surface.}
  \label{figure:plotfes}
\end{figure}

Typing the name of the variable with a free energy surface returns its dimensionality,
number of points, and free energy maximum and minimum. The same is returned by
\code{summary} function. It is possible
to add and subtract two free energy surfaces with the same number of grid points.
The functions \code{min} and \code{max} can be used as well to calculate minimum or
maximum. It is also possible to multiply or divide the free
energy surface by a constant (for example, to convert kJ to kcal and vice versa).
Free energy surface can be plotted (Figure \ref{figure:plotfes}) by typing:
\begin{example}
  plot(fesurface, xlab="phi", ylab="psi")
\end{example}

In metadynamics simulation, it is important to find free energy minima. The global minimum
refers to the most favored state of the system (i.e., the state with the highest probability).
Other local minima correspond to metastable states. The user can find free energy
minima by typing:
\begin{example}
  minima <- fesminima(fesurface)
\end{example}
This function locates minima using a simple algorithm. The free energy surface is separated into
8, 8x8, or 8x8x8 bins (for 1D, 2D, or 3D surface, respectively). A minimum in each bin is
located. Next, the program tests whether a minimum is a local minimum of the whole free
energy surface. The number of grid points can be changed by \code{ngrid} parameter.
Typing the name of the minima variable will return the table of minima (denoted as
A, B, C, ... in the order of their free energies), their collective variables, and free
energy values.

In addition, the function summary provides populations of each
minimum calculated as:
\begin{equation}
P_{i,rel} = \exp(-G_i/kT)
\end{equation}
\begin{equation}
P_i = P_{i,rel} / \sum (P_{j,rel})
\end{equation}
\begin{example*}
  letter CV1bin CV2bin        CV1        CV2 free_energy relative_pop
1      A     78    236 -1.2443171  2.6487938   -97.26095 8.614856e+16
2      B     28    240 -2.4763142  2.7473536   -95.63038 4.480527e+16
3      C     74    118 -1.3428769 -0.2587194   -94.73163 3.124915e+16
4      D    166    151  0.9239978  0.5543987   -91.66626 9.143024e+15
5      E    170    251  1.0225576  3.0183929   -84.37799 4.920882e+14
         pop
1 50.1335658
2 26.0741201
3 18.1852268
4  5.3207200
5  0.2863674
\end{example*}

\begin{figure}[htbp]
  \centering
  \includegraphics{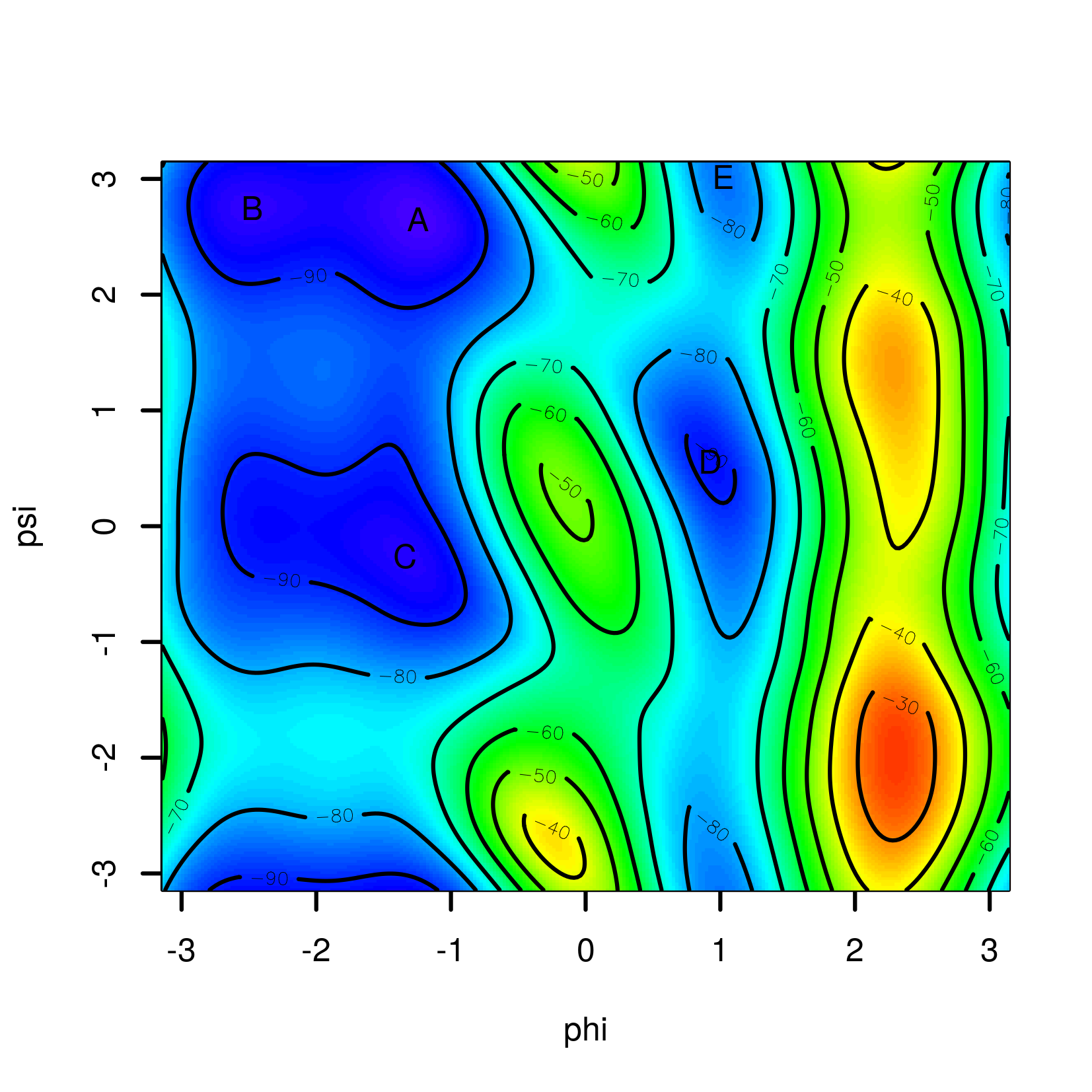}
  \caption{Free energy surface with indicated free energy
           minima A-E.}
  \label{figure:minima}
\end{figure}

Plot function on a \code{fesminima} output provides the same plot
as for \code{fes} output with additional letters
indicating minima (Figure~\ref{figure:minima}).

\begin{figure}[htbp]
  \centering
  \includegraphics{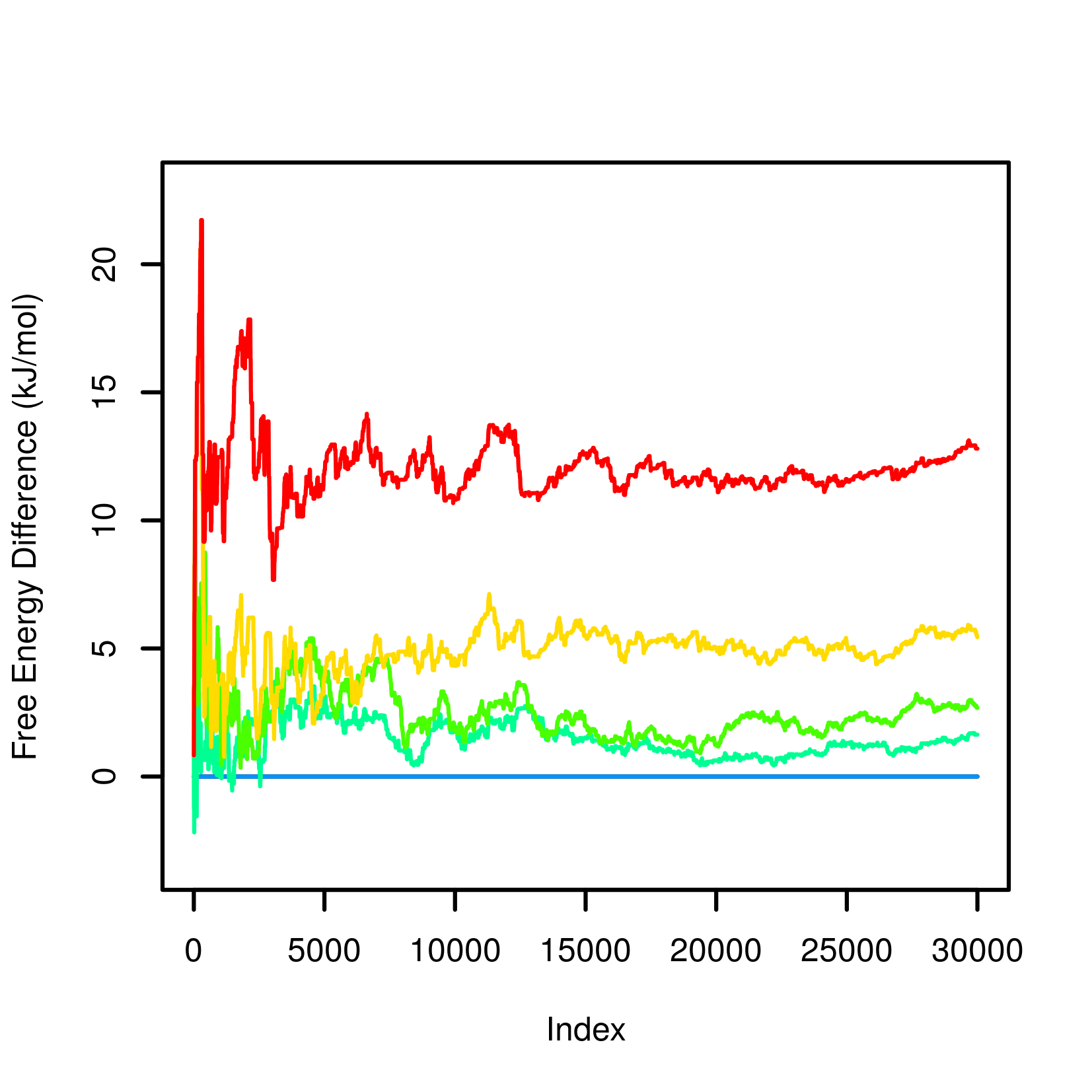}
  \caption{Evolution of free energy differences.}
  \label{figure:conv}
\end{figure}

It is essential to evaluate the accuracy of metadynamics and to decide
when the simulation is accurate enough so that it can be stopped. For this
purpose, it is useful to look at the evolution of relative free energies.
The relative free energies (for example, the free energy difference of minima
A and C) evolve rapidly at the beginning of the simulation, and with the progress
of the simulation, their difference is converging towards the real free energy
difference. Function \code{feprof} calculates the evolution of free energy differences
from the global minimum (global at the end of the simulation).
It can be used as:
\begin{example}
  prof<-feprof(minima)
\end{example}
Function summary provides minima and maxima of these free energy differences. The evolution
can be plotted (Figure \ref{figure:conv}) by typing:
\begin{example}
  plot(prof)
\end{example}

Beside minima, another important points on the free energy surface are transition
states. Change of the molecular structure from one minimum to
another takes place via a path with the lowest energy demand.
The state with the highest energy along this path is called
the transition state. Free energy difference between the initial
and transition state can be used to predict kinetics (rates)
of the studied molecular process. Furthermore, identification
of transition states is important in drug design because
compounds designed to mimic a transition states of an enzymatic
reaction are often potent enzymes inhibitors and
drugs \citep{relenza}.

\begin{figure}[htbp]
  \centering
  \includegraphics{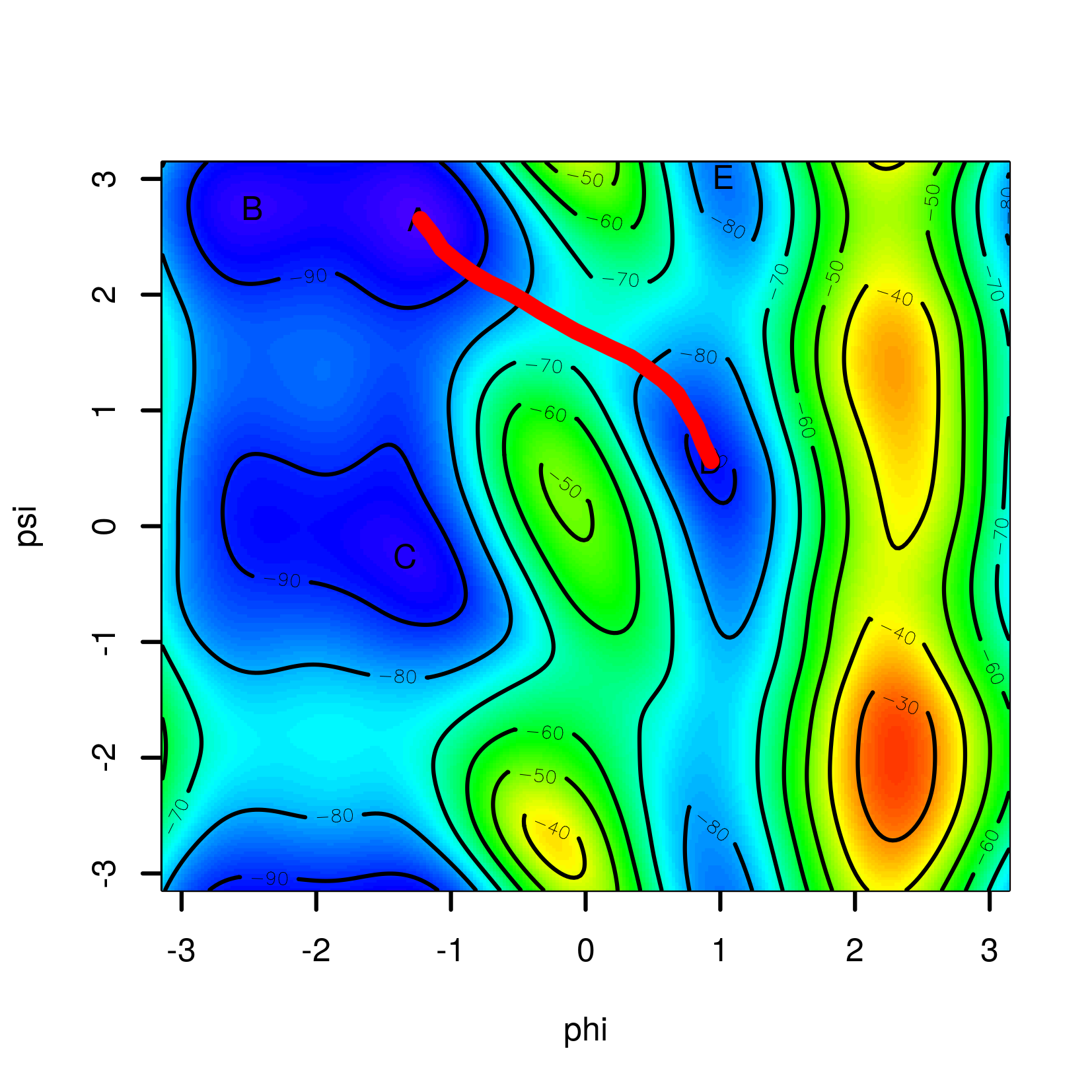}
  \caption{Path of transition from A to D projected onto
           free energy surface.}
  \label{figure:neb}
\end{figure}

In \pkg{metadynminer}, such path can be identified by Nudged
Elastic Band method \citep{neb}. Briefly, this method plots
a line between selected minima as an initial approximation
of the transition path. Next, this line is curved so that the
corresponding physical process becomes feasible. This function
can be applied on, for example, minima A and D as:
\begin{example}
  nebAD<-neb(minima, min1="A", min2="D")
\end{example}
The result can be analyzed by \code{summary} (to provide kinetics
of the A to D and D to A change predicted by Eyring
equation \citep{eyring}), by \code{plot} (to plot the free energy
profile of the molecular process) and by \code{pointsonfes}
of \code{linesonfes} (to plot the path on top of the free energy
surface). The last example can be invoked by:
\begin{example}
  plot(minima, xlab="phi", ylab="psi")
  linesonfes(nebAD, lwd=4)
\end{example}
The resulting plot is depicted in Figure~\ref{figure:neb}

\begin{figure}[htbp]
  \centering
  \includegraphics{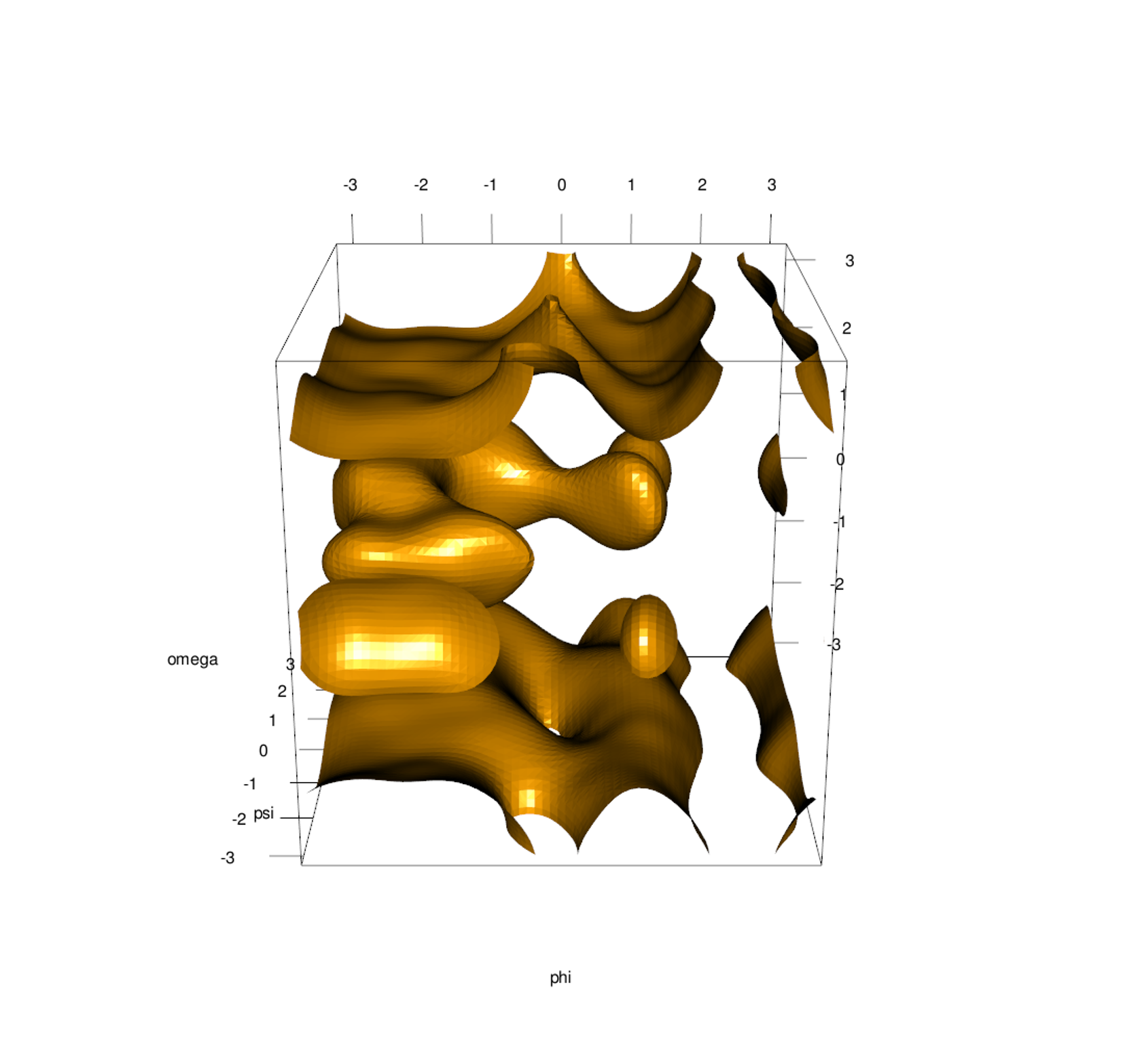}
  \caption{3D free energy surface depicted as isosurface at $-30$ kJ/mol.}
  \label{figure:fes3d}
\end{figure}

Let us also briefly present \pkg{metadynminer3d}. This package uses
packages \CRANpkg{rgl} and \CRANpkg{misc3d} to plot the free energy surface as
an interactive (mouse rotatable) isosurface in the space of three collective variables (see
Figure \ref{figure:fes3d}). \pkg{Metadynminer3d} can produce interactive WebGL visualizations
using \code{writeWebGL} command from the \pkg{rgl} package.

\pkg{Metadynminer} and \pkg{metadynminer3d} were developed to be highly flexible.
This flexibility can be demonstrated on two examples. First, it is
useful to visualize the progress of metadynamics as a video sequence showing
the evolution of the free energy surface. The code to generate
corresponding images can be written in \pkg{metadynminer} as:
\begin{example}
tfes<-fes(acealanme, tmax=100)
png("snap
plot(acealanme, zlim=c(-200,0))
for(i in 1:299) {
  tfes<-tfes+fes(acealanme, imin=100*i+1, imax=100*(i+1))
  plot(tfes, zlim=c(-200,0), xlab="phi", ylab="psi")
}
dev.off()
\end{example}
This generates a series of images that can be concatenated by external
software to make a video file.

\begin{figure}[htbp]
  \centering
  \includegraphics{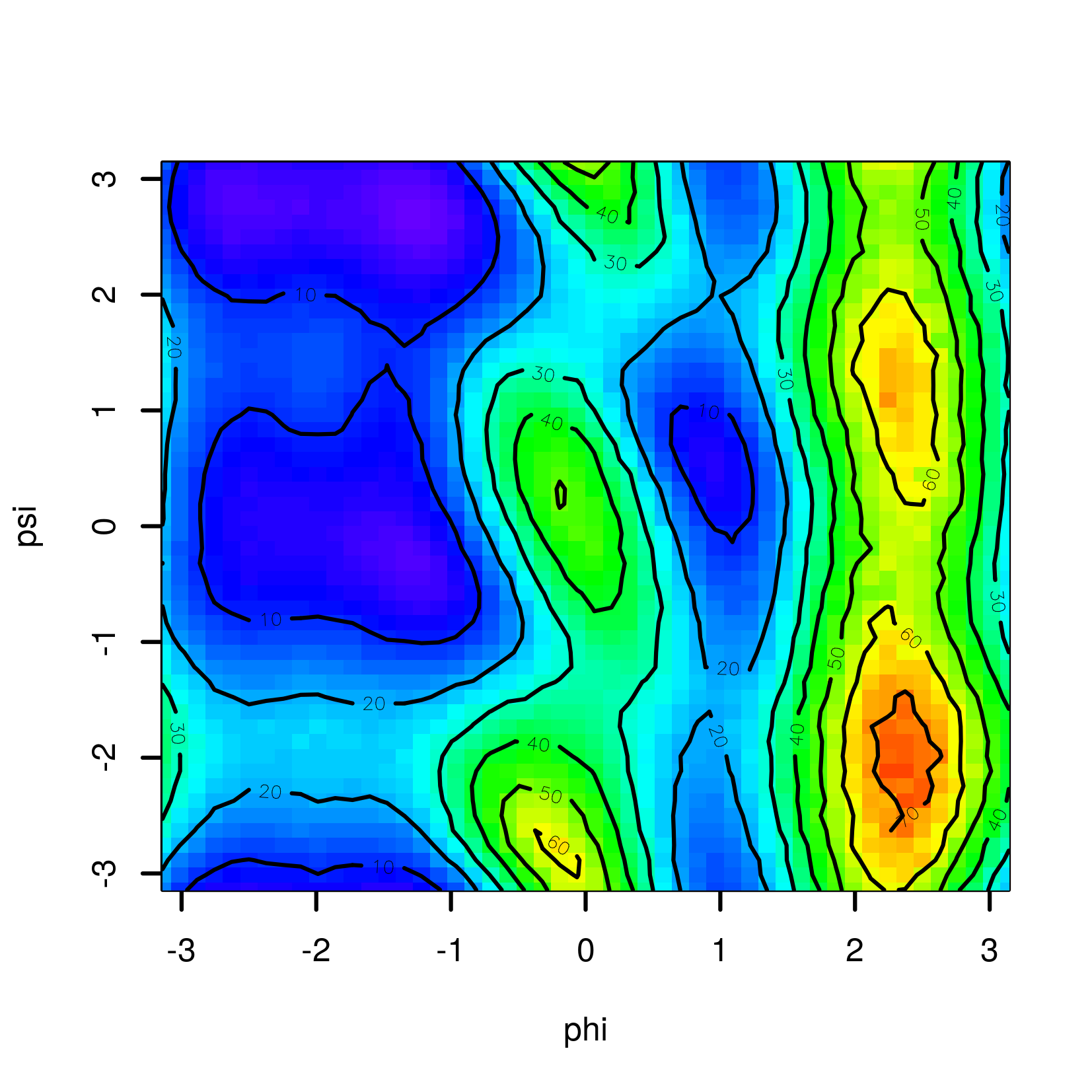}
  \caption{Free energy surface calculated by reweighting by
           \cite{tiwary}.}
  \label{figure:rew}
\end{figure}

The second example demonstrates a more complicated analysis of the
results from metadynamics. Functions \code{fes} and \code{fes2} use equations
\ref{mtd} and \ref{wtmtd} to predict the free energy surface. A limitation
of this approach is that the prediction of the free energy surface is
based only on the positions of hills. The evolution of collective variables between
hills depositions is not used. As an alternative, it is possible to use
reweighting \citep{us,tiwary}. This approach calculates the free
energy surface from hills positions as well as from evolution of collective
variables. Briefly, regions of the free energy surface that are sampled
despite being disfavored by high flooding potential have higher weights than those
disfavored by low flooding potential. This approach is in general more accurate.
Reweighting can be done using the code:
\begin{example*}
bf <- 15
kT <- 8.314*300/1000
npoints <- 50
maxfes <- 75
outfes <- 0*fes(acealanme, npoints=npoints)
s1 <- c()
s2 <- c()
for(i in 1:50) \{
  step <- i*length(acealanme\$time)/50
  cfes <- fes(acealanme, imax=step)
  s1 <- c(s1, sum(exp(-cfes\$fes/kT)))
  s2 <- c(s2, sum(exp(-cfes\$fes/kT/bf)))
\}
ebetac <- s1/s2
cvs <- read.table("COLVAR")
nsamples <- nrow(cvs)
xlim <- c(-pi,pi)
ylim <- c(-pi,pi)
for(i in 1:nsamples) \{
  step <- (i-1)*50/nsamples+1
  ix <- npoints*(cvs[i,2]-xlim[1])/(xlim[2]-xlim[1])+1
  iy <- npoints*(cvs[i,3]-ylim[1])/(ylim[2]-ylim[1])+1
  outfes\$fes[ix,iy] <- outfes\$fes[ix,iy] + exp(cvs[i,4]/kT)/ebetac[step]
\}
outfes\$fes[!outfes\$fes>0] <- maxfes
outfes\$fes <- -kT*log(outfes\$fes)
plot(outfes, xlab="phi", ylab="psi")
\end{example*}
where \code{bf} is the bias factor ($(T+\Delta T)/T$ in Equation \ref{wtmtd}),
\code{kT} is temperature in Kelvins multiplied by Boltzmann constant,
\code{npoints} is the granularity of the resulting free energy surface
and \code{maxfes} is the maximal possible free energy (to avoid problems
with infinite free energy in unsampled regions).
First, \code{outfes} is introduced as a zero
free energy surface. The first loop calculates the correction \code{ebetac} for the
evolution of flooding potential developed by Tiwary and Parrinello
\citep{tiwary}. Next, a file with the evolution of collective variables
\code{COLVAR} (from the same simulation used to generate \code{acealanme}
dataset, available at \url{https://www.metadynamics.cz/metadynminer/data/})
is read. The second loop evaluates the sampling weighted by the factor
$\exp (V(s)/kT)$ divided by \code{ebetac} to correct for the evolution of
the bias potential \citep{tiwary}. Finally, probabilities are converted
to the free energy surface and plotted (Figure \ref{figure:rew}).

\section{Simulation details} 
All simulations were done using Gromacs 2016.4 \citep{gmx} patched by
Plumed 2.4b \citep{plumed}. Alanine dipeptide was modeled using
Amber99SB-ILDN force field \citep{ildn}. The simulated system
contained alanine dipeptide and 874 TIP3P \citep{tip3p} water
molecules. The temperature was kept constant at 300 K using Bussi
thermostat \citep{vrescale}. Metadynamics hills of height
1 kJ/mol (bias factor 10) and widths 0.3 rad were added
every 1 ps. Two simulations were done, one with
one dihedral angle $\phi$ (dataset \code{acealanme1d}),
two dihedral angles $\phi$ and $\psi$ (dataset \code{acealanme}),
or with three angle $\phi$, $\phi$ and $\omega$ (dataset
\code{acealanme3d} in \pkg{metadynminer3d}).
Supporting material is available at
\url{https://www.metadynamics.cz/metadynminer/data/}
or in Plumed nest \citep{nest} at
\url{https://www.plumed-nest.org/eggs/20/023/}.

\section{Summary}
The package \pkg{metadynminer} and \pkg{metadynminer3d} provides fast algorithm
Bias Sum \citep{petr} for calculation of free energy surfaces from
metadynamics. This algorithm is available in our on-line tool
MetadynView (\url{http://metadyn.vscht.cz}), but this tools is
intended for routine checking of the progress of metadynamics
simulations rather than for in deep analysis and visualization.
Besides this, users of metadynamics use built-in functions in
Plumed or various in-lab scripts. Such scripts do not provide
appropriate flexibility in analysis and visualization.

We see the biggest advantage in the fact that 
\pkg{metadynminer} can produce publication quality figures
via graphics output functions in R. As shown above,
using a simple \code{for} loop it is possible to plot
individual snapshots and concatenate them outside R
to make a movie. \pkg{Metadynminer3d} provides the possibility
to produce interactive 3D web models by WebGL technology.
We also tested 3D printing of a free energy surface that
is very easy using \pkg{metadynminer} and \pkg{rayshader}. 
Various tips and tricks can be found
on the website of the project
(\url{https://www.metadynamics.cz/metadynminer/}).

Another advantage we see in reporting of results. Reproducibility
is a big issue in science, including molecular simulations.
Packages like \CRANpkg{knitr} or \CRANpkg{rmarkdown} ca be
used to record all steps of data analysis pipeline to compile
a report for routine and reproducible use of metadynamics.

\section{Acknowledgement}

This project was supported by Ministry of Education, Youth and
Sports of the Czech Republic - COST action OpenMultiMed 
(CA15120, LTC18074) for development and Czech National
Infrastructure for Biological data (ELIXIR CZ, LM2015047)
for future sustainability.

\bibliography{RJreferences}

\address{Dalibor Trapl\\
  Department of Biochemistry and Microbiology, University of Chemistry and Technology, Prague\\
  Technicka 3, Prague 6, 166 28\\
  Czech Republic\\
  ORCID: 0000-0002-3435-5841\\
  \email{traplda@vscht.cz}}

\address{Vojtech Spiwok\\
  Department of Biochemistry and Microbiology, University of Chemistry and Technology, Prague\\
  Technicka 3, Prague 6, 166 28\\
  Czech Republic\\
  ORCID: 0000-0001-8108-2033\\
  \email{spiwokv@vscht.cz}}

\end{article}

\end{document}